%% file: main.tex
\documentclass[11pt]{journal}

\usepackage[margin=0.7in]{geometry}
\usepackage{graphicx}
\usepackage{amssymb}
\usepackage{float}
\usepackage{authblk}
\usepackage{natbib}
\usepackage[rgb]{xcolor}
\usepackage{mathsymbols}
\usepackage[cmex10]{amsmath}
\usepackage{hyperref}
\usepackage{rotating}
\usepackage{multirow}
\usepackage[textwidth=2.450cm, textsize=tiny]{todonotes}
\usepackage{enumitem}
\usepackage[]{helvet}
\usepackage[font=small,labelfont=bf]{caption}

\hypersetup{
    colorlinks,
    citecolor=blue
}

\newcommand{\overbar}[1]{\mkern 1.5mu\overline{\mkern-1.5mu#1\mkern-1.5mu}\mkern 1.5mu}

\author[1]{\normalsize Mainak Jas \footnote{46 Rue Barrault, T\'el\'ecom ParisTech, Universit\'e Paris-Saclay, France. E-mail address: \href{mailto:mainak.jas@telecom-paristech.fr}{mainak.jas@telecom-paristech.fr}}}
\author[2, 3, 7]{Denis A. Engemann\protect\footnotemark\protect\footnotetext[2]{equal contributions}}
\author[1]{Yousra Bekhti}
\author[4, 5, 6, 7]{Federico Raimondo}
\author[1]{Alexandre Gramfort\protect\footnotemark[2]}

\affil[1]{LTCI, T\'el\'ecom ParisTech, Universit\'e Paris-Saclay, France}
\affil[2]{Parietal project-team, INRIA Saclay - ile de France, France}
\affil[3]{Cognitive Neuroimaging Unit, CEA DSV/I2BM, INSERM, Universit{\'e} Paris-Sud,\\ Universit{\'e} Paris-Saclay, NeuroSpin center, 91191 Gif/Yvette, France}
\affil[4]{Laboratorio de Inteligencia Artificial Aplicada, Departamento de Computaci\'on, FCEyN, Universidad de Buenos Aires, Argentina}
\affil[5]{CONICET, Argentina}
\affil[6]{Sorbonne Universit\'es, UPMC Univ Paris 06, Facult\'e de M\'edecine Piti\'e-Salp\^etri\`ere, Paris, France}
\affil[7]{Institut du Cerveau et de la Moelle \'epini\`ere, ICM, PICNIC Lab, F-75013, Paris, France}

\title{Autoreject: Automated artifact rejection for MEG and EEG data}

\begin{document}

\maketitle

\input{sections/abstract.tex}
\input{sections/introduction.tex}
\input{sections/methods.tex}
\input{sections/validation_protocol.tex}

\input{sections/results.tex}

\input{sections/discussion.tex}

\input{sections/conclusion.tex}
\input{sections/acknowledgement.tex}

\bibliographystyle{model1-num-names}
\bibliography{refs.bib}

\input{sections/supplementary.tex}

\end{document}

%% file: sections/abstract.tex
\begin{abstract}
We present an automated algorithm for unified rejection and repair of bad trials in magnetoencephalography (MEG) and electroencephalography (EEG) signals. Our method capitalizes on cross-validation in conjunction with a robust evaluation metric to estimate the optimal peak-to-peak threshold -- a quantity commonly used for identifying bad trials in M/EEG. This approach is then extended to a more sophisticated algorithm which estimates this threshold for each sensor yielding trial-wise bad sensors. Depending on the number of bad sensors, the trial is then repaired by interpolation or by excluding it from subsequent analysis. All steps of the algorithm are fully automated thus lending itself to the name \textit{Autoreject}.

In order to assess the practical significance of the algorithm, we conducted extensive validation and comparisons with state-of-the-art methods on four public datasets containing MEG and EEG recordings from more than 200 subjects. The comparisons include purely qualitative efforts as well as quantitatively benchmarking against human supervised and semi-automated preprocessing pipelines. The algorithm allowed us to automate the preprocessing of MEG data from the Human Connectome Project (HCP) going up to the computation of the evoked responses. The automated nature of our method minimizes the burden of human inspection, hence supporting scalability and reliability demanded by data analysis in modern neuroscience.
\end{abstract}

\begin{keywords}
Magnetoencephalography (MEG), Electroencephalogram (EEG), preprocessing, statistical learning, cross-validation, automated analysis, Human Connectome Project (HCP)

\end{keywords}

%% file: sections/introduction.tex
\section{Introduction}
\label{sec:introduction}

\begin{figure}[t]
	\centering
	\includegraphics[width=0.55\linewidth]{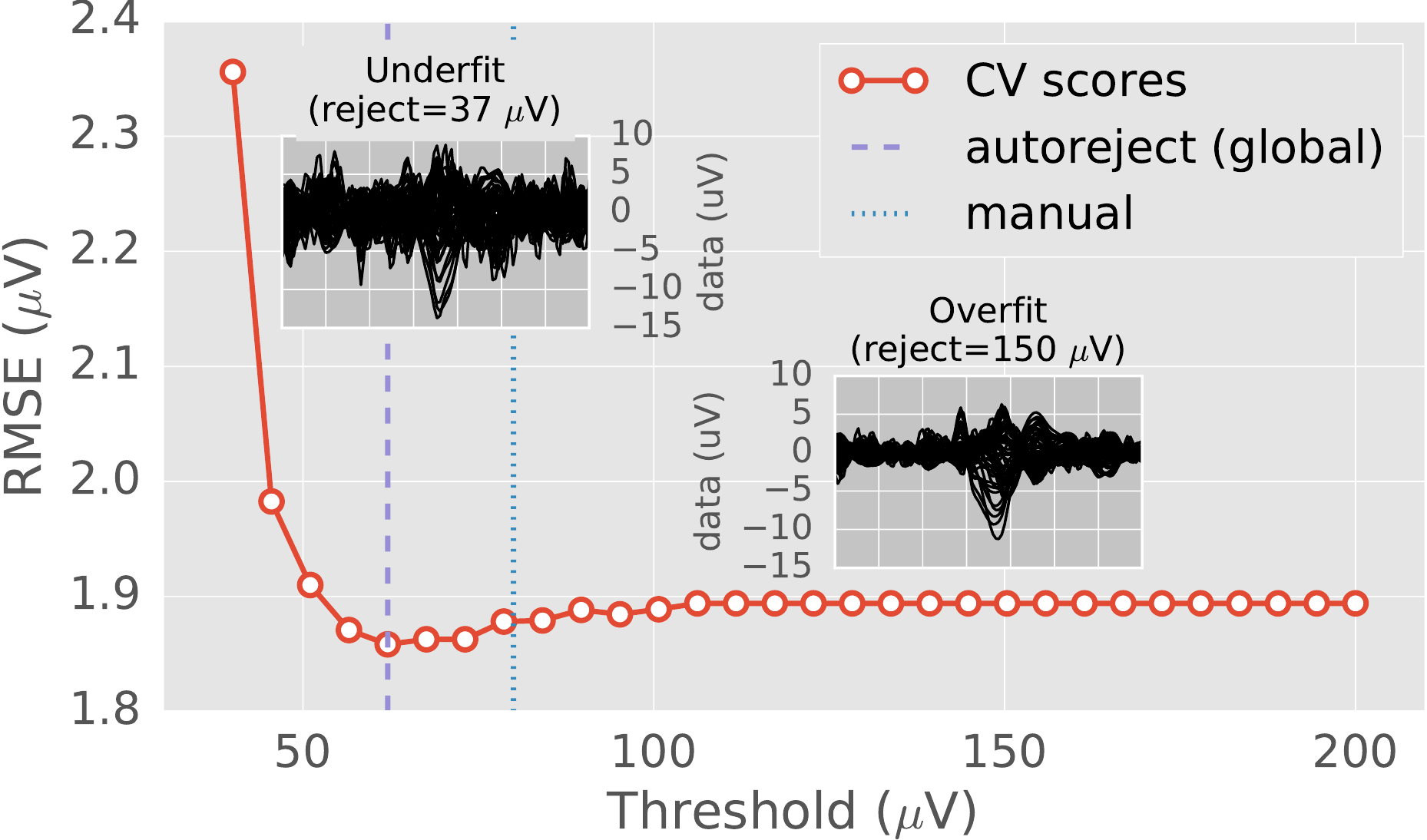}
    \caption{Cross-validation error as a function of peak-to-peak rejection threshold on one EEG dataset. The root mean squared error (RMSE) between the mean of the training set (after removing the trials marked as bad) and the median of the validation set was used as the cross-validation metric (Section~\ref{sec:auto_global}). The two insets show the average of the trials as ``butterfly plots" (each curve representing one sensor) for very low and high thresholds. For low thresholds, the RMSE is high because most of the trials are rejected (underfit). At high thresholds, the model does not drop any trials (overfit). The optimal data-driven threshold (\emph{autoreject, global}) with minimum RMSE is somewhere in between. It closely matches the human threshold.}
    \label{fig:cross_val}
\end{figure}

Magneto-/electroencephalography (M/EEG) offer the unique ability to explore
and study, non-invasively, the temporal dynamics of the brain and its cognitive processes. The M/EEG community has only recently begun to appreciate the importance of large-scale studies, in an effort to improve replicability and statistical power of experiments. This has given rise to the practice of sharing and publishing data in open archives~\citep{gorgolewski2016practical}. Examples of such large electrophysiological datasets include the Human Connectome Project (HCP)~\citep{van2012human, larson2013adding}, the Physiobank~\citep{goldberger2000physiobank}, the OMEGA archive~\citep{niso2016omega} and Cam-CAN~\citep{taylor2015cambridge}. A tendency towards ever-growing massive datasets as well as a shift towards common standards for accessing these databases~\citep{gorgolewski2016brain,bigdely2013hierarchical} is clearly visible. The UK Biobank project~\citep{ollier2005uk} which currently hosts data from more than 50,000 subjects is yet another example of this trend.

This has however, given rise to new challenges including automating the analysis pipeline~\citep{gorgolewski2016practical}. Automation will not only save time, but also allow scalable analysis and reduce the barriers to reanalysis of data, thus facilitating reproducibility. \citet{engemann2015automated_new} have recently worked towards more automation in M/EEG analysis pipelines by considering the problem of covariance estimation, a step commonly done prior to source localization. Yet, one of the most critical bottlenecks that limits the reanalysis of M/EEG data remains at the preprocessing stage with the annotation and rejection of artifacts. Despite being so fundamental to M/EEG analysis given how easily such data can be corrupted by noise and artifacts, there is currently no consensus in the community on how to address this particular issue.

In the presence of what we will refer to as \emph{bad} data, various data cleaning strategies have been employed. A first intuitive strategy is to exclude bad data from analysis, to \emph{reject} it. While this approach is very often employed, for example, because data cleaning is time consuming, or out
of reach for practitioners, it leads to a loss of data that are costly to acquire. This is particularly the case for clinical studies, where patients have difficulties staying still or focusing on the task~\citep{cruse2012bedside,goldfine2013reanalysis}, or even when babies are involved as subjects~\citep{basirat2014hierarchy}.

When working with M/EEG, the data can be bad due to the presence of bad sensors (also known as channels\footnote{They are not necessarily equivalent in the case of a bipolar montage in EEG. However, for the sake of simplicity, we shall use these terms interchangeably in this work.}) and bad trials.
A trial refers here to a data segment whose location in time is typically related to an experimental protocol. But here we will also call trial any data segment even if it is acquired during a task-free protocol.
Accordingly, a bad trial or bad sensor is one which contains bad data.
Ignoring the presence of bad data can adversely affect analysis downstream in the pipeline. For example, when multiple trials time-locked to the stimulation are averaged to estimate an evoked response,
ignoring the presence of a single bad trial can corrupt the average. The mean of a random vector is not robust to the presence of strong outliers. Another example quite common in practice, both in the case of EEG and MEG, is the presence of a bad sensor. When kept in the analysis, an artifact present on a single bad sensor can spread to other sensors, for example due to spatial projection. This is why identifying bad sensors is crucial for data cleaning techniques such as the very popular Signal Space Separation (SSS) method~\citep{taulu2004suppression}. Frequency filtering~\citep{widmann2015digital} can often suppress many low frequency artifacts, but turns out to be insufficient for broadband artifacts. A common practice to mitigate this issue is to visually inspect the data using an interactive viewer and mark manually, the bad sensors and bad segments in the data. Although trained experts are very likely to agree on the annotation of bad data, their judgement is subject to fluctuations and cannot be repeated. Their judgement can also be biased due to prior training with different experimental setups or equipments, not to mention the difficulty for such experts to allocate some time to review the raw data collected everyday.

Luckily, popular software tools such as
Brainstorm~\citep{tadel2011brainstorm}, 
EEGLAB~\citep{delorme2004eeglab}, 
FieldTrip~\citep{oostenveld2010fieldtrip},
MNE~\citep{gramfort2013meg}
or SPM~\citep{litvak2011eeg} already allow for the rejection of bad data segments based on simple metrics such as peak-to-peak signal amplitude differences that are compared to a manually set threshold value. When the peak-to-peak amplitude in the data exceeds a certain threshold, it is considered as bad. However, while this seems quite easy to understand and simple to use from a practitioner's standpoint, this is not always convenient. In fact, a good peak-to-peak signal amplitude threshold turns out to be data specific, which means that setting it requires some amount of trial and error.

The need for better automated methods for data preprocessing is clearly shared by various research teams, as the literature of the last few years can confirm. On the one hand, are pipeline-based approaches, such as Fully Automated Statistical Thresholding for EEG artifact rejection (FASTER by~\citet{nolan2010faster}) which detect bad sensors as well as bad trials using fixed thresholds motivated from classical Gaussian statistics. Methods such as PREP~\citep{bigdely2015prep}, on the other hand, aim to detect and clean the bad sensors only. Unfortunately, they do not offer any solution to reject bad trials. Other methods are available to solve this problem. For example, the Riemannian Potato~\citep{barachant2013riemannian} technique can identify the bad trials as those where the covariance matrix lies outside of the ``potato'' of covariance matrices for good trials. By doing so, it marks trials as bad but does not identify the sensors causing the problem, hence not offering the ability to repair them. It appears that practitioners are left to choose between different methods to reject trials or repair sensors, whereas they are in fact intricately related problems and must be dealt with together. 

Robust regression~\citep{diedrichsen2005detecting} also deals with bad trials using a weighted average which mitigates the effect of outlier trials. Trials with artifacts end up with low contributions in the average. A related approach that is sometimes employed to ignore outlier trials in the average is the trimmed mean as opposed to a regular mean. The trimmed mean is a compromise between the mean which offers a high signal-to-noise ratio (SNR) but can be corrupted by outliers, and the median which is immune to outliers of extreme amplitudes but has a low SNR as it involves no averaging. Of course, neither of these strategies are useful when analyses have to be conducted on single trials. Another approach, which is also data-driven, is Sensor Noise Suppression (SNS)~\citep{de2008sensor}. It removes the sensor-level noise by spatially projecting the data of each sensor onto the subspace spanned by the principal components of all the other sensors. This projection is repeated in leave-one-sensor-out iterations so as to eventually clean all the sensors. In most of these methods, however, there are parameters which are somewhat dataset dependent and must therefore be manually tuned.

We therefore face the same problem in automated methods as in the case of semi-automated methods such as peak-to-peak rejection thresholds, namely the tuning of model parameters. In fact, setting the model parameters is even more challenging in some of the methods when they do not directly translate into human-interpretable physical units.

This led us to adopt a pragmatic approach in terms of algorithm design, as it focuses on the tuning of the parameters that M/EEG users presently choose manually. The goal is, not only to obtain high quality data but also to develop a method which is transparent and not too disruptive for the majority of M/EEG users. A first question we address below is: can we improve peak-to-peak based rejection methods by automating the process of trial and error? In the following section, we explain how the widely-known statistical method of cross-validation (see Figure~\ref{fig:cross_val} for a preview) in combination with Bayesian optimization~\citep{snoek2012practical, bergstra2011algorithms} can be employed to tackle the problem at hand. We then explain how this strategy can be extended to set thresholds separately for each sensor and mark trials as bad when a large majority of the sensors have high-amplitude artifacts. This process closely mimics how a human expert would mark a trial as bad during visual inspection. 

In the rest of the paper, we detail the internals of our algorithm, compare it against various state-of-the-art methods, and position it conceptually with respect to these different approaches. For this purpose, we make use of qualitative visualization techniques as well as quantitative reports. In a major validation effort, we take advantage of cleaned up evoked response fields (ERFs) provided by the Human Connectome Project~\citep{larson2013adding} enabling ground truth comparison between alternative methods. This work represents one of the first efforts in reanalysis of the MEG data from the HCP dataset using a toolkit stack significantly different from the one employed by the HCP consortium. The convergence between our method and the HCP MEG pipelines is encouraging and testifies to the success of the community-wide open science efforts aiming at reproducible research. Naturally, we have therefore made our code available online\footnote{\url{https://autoreject.github.io}}. In addition to this, we validated our algorithm on the MNE sample data~\citep{gramfort2013meg}, the multimodal faces dataset~\citep{wakeman2015multi}, and the EEGBCI motor imagery data~\citep{goldberger2000physiobank,schalk2004bci2000}.

A preliminary version of this work was presented in~\citet{jas2016automated}.
\paragraph{Notations} We denote matrices by capital letters $X \in \real^{m \times n}$. The $i$th row of a matrix is indexed by subscripts, as in $X_{i}$, and the entry in the $i$th row and $j$th column is indexed as $X_{ij}$. The matrix $X$ restricted to the rows with indices in the set $\mathcal{G}$ is denoted by $X_\mathcal{G}$. All sets $\mathcal{G}$, $\mathcal{T}$ or $\mathcal{V}$ are written in calligraphic fonts.

%% file: sections/methods.tex
\section{Materials and methods}
We will first describe how a cross-validation procedure can be used to set peak-to-peak rejection thresholds globally (\textit{i.e.} same threshold for all sensors). This is what we call \textit{autoreject (global)}.

\subsection{Autoreject (global)}
\label{sec:auto_global}
We denote the data matrix by $X \in \real^{N \times P}$ with $N$ trials and $P$ features. These $P$ features are the $Q$ sensor-level time series, each of length $T$ concatenated along the second dimension of the data matrix, such that $P=QT$. We divide the data into $K$ folds (along the first dimension) with training set indices $\mathcal{T}_{k}$ and validation set indices $\mathcal{V}_{k}=[1..N] \setminus {\mathcal{T}_k}$ for each fold $k$ $(1 \leq k \leq K)$. For simplicity of notation, we first define the peak-to-peak amplitude for the $i$th trial and $j$th sensor as the difference between the maximum and the minimum value in that time series:
\begin{equation}
\mathcal{A}_{ij} = \max_{(j-1)T+1 \leq t \leq jT} (X_{it}) - \min_{(j-1)T+1 \leq t \leq jT} (X_{it}) \enspace .
\end{equation}
The good trials $\mathcal{G}_k$ where the peak-to-peak amplitude $\mathcal{A}_{ij}$ for any sensor does not exceed the candidate threshold $\tau$ are computed as
\begin{equation}
\mathcal{G}_k = \{i \in \mathcal{T}_{k} \suchthat \max_{1 \leq j \leq Q} \mathcal{A}_{ij} \leq \tau\}.
\end{equation}
By comparing the peak-to-peak threshold with the maximum of the peak-to-peak amplitudes, we ensure that none of the sensors exceed the given threshold. Once we have applied the threshold on the training set, it is necessary to evaluate how the threshold performs by looking at new data. For this purpose, we consider the validation set. Concretely speaking, we propose to compare the mean $\overbar{X_{\mathcal{G}_k}}$ of good trials in the training set against the median $\widetilde{X_{\mathcal{V}_k}}$ of all trials in the validation set. Using root mean squared error (RMSE) the mismatch $e_{k}(\tau)$ reads as:
\begin{equation}
 e_{k}(\tau) = \fro{\overbar{X_{\mathcal{G}_k}} - \widetilde{X_{\mathcal{V}_k}}}.
\label{eq:err} 
\end{equation}
Here, $\fro{\cdot}$ is the Frobenius norm. The rationale for using the median in the validation set is that it is robust to outliers. Indeed, it is far less affected by high-amplitude artifacts than a classical mean. The threshold with the best data quality (lowest mismatch $e_{k}(\tau)$) on average across the $K$ folds is selected as the optimal threshold. In practice $\tau$ is taken in a bounded interval $[\tau_{\min}, \tau_{\max}]$:
\begin{equation}
\tau_{\star} = \underset{\tau \in [\tau_{\min}, \tau_{\max}]} \argmin \frac{1}{K} \sum_{k=1}^{K}  e_{k}(\tau)
\label{eq:best_th}
\end{equation}
Note, that $\widetilde{X_{\mathcal{V}_k}}$ does not depend on $\tau$. Indeed, it would not be wise to restrict the validation set to good trials according to the value of $\tau$. As $\tau$ varies, it would lead to a variable number of validation trials, which would affect the comparison of RMSE across threshold values. The idea of using the median in the context of cross-validation has been previously proposed in the statistics literature in order to deal also with outliers~\citep{leung2005cross,de2003robust}.

Figure~\ref{fig:cross_val} shows how the average RMSE changes as the threshold varies for the MNE sample dataset~\citep{gramfort2013meg,mne}. At low thresholds, our model underfits as it drops most of the trials in the data resulting in a noisy average. On the other hand, at high thresholds, the model overfits retaining all the trials in the data including the high-amplitude artifacts. Here the candidate values of $\tau$ were taken on a grid. More details on how to solve \eqref{eq:best_th} will be given in Section~\ref{sec:bayesian_opt}.

\subsection{Autoreject (local)}
\label{sec:auto_local}

\begin{figure}[t]
	\centering
	\includegraphics[width=0.5\linewidth]{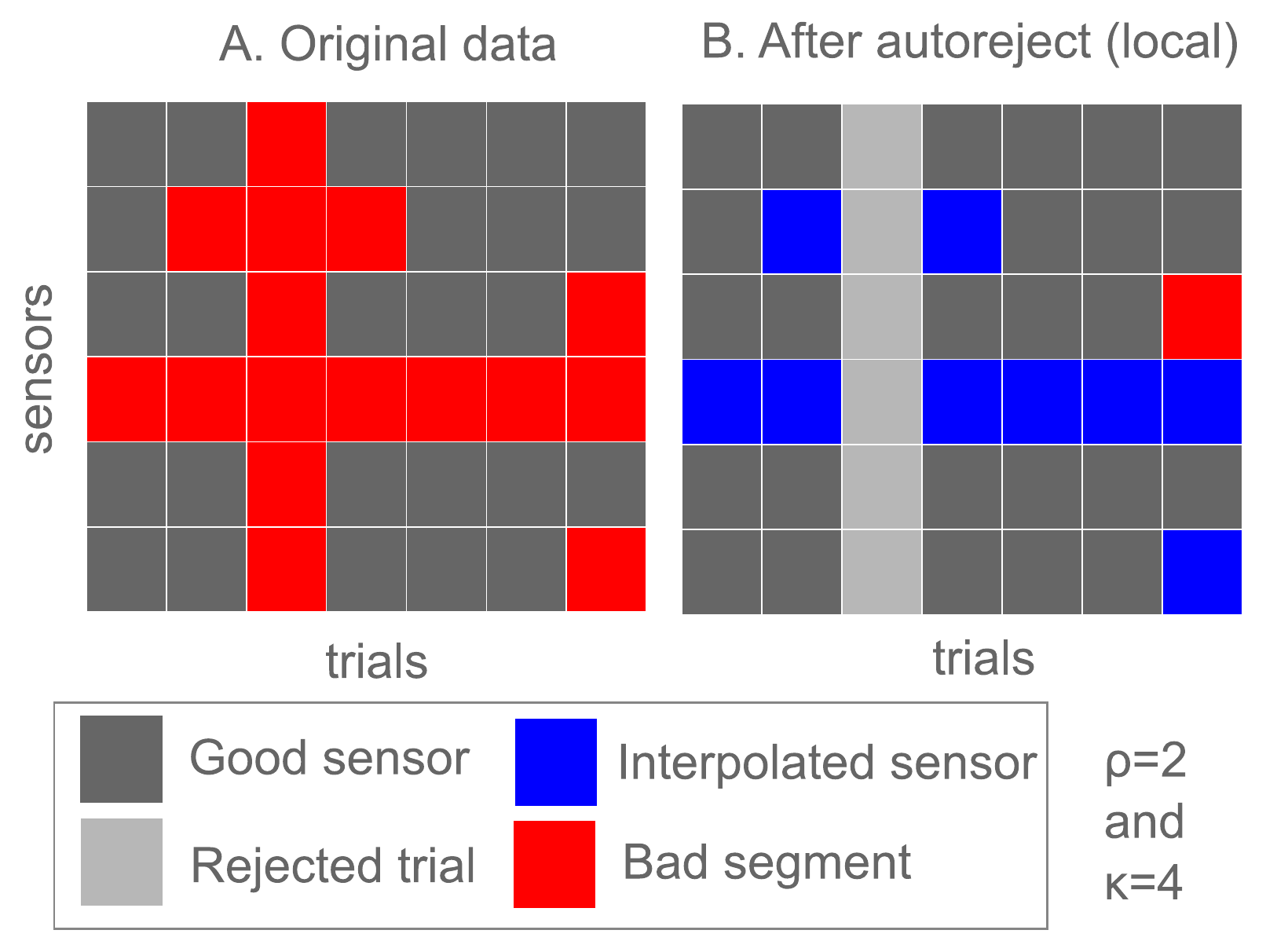}
    \caption{A schematic diagram explaining how \emph{autoreject (local)} works. (A) Each cell here is an element of the indicator matrix $C_{ij}$ described in Section~\ref{sec:auto_local}. Sensor-level thresholds are found and bad segments are marked for each sensor. Bad segments shown in red are where $C_{ij}=1$ (B) Trials are rejected if the number of bad sensors is greater than $\kappa$ and otherwise, the worst $\rho$ sensors are interpolated.}
    \label{fig:schematic}
\end{figure}

A global threshold common to all sensors, however, suffers from limitations. A common case of failure is when a single sensor is affected (locally or globally) by high-amplitude artifacts. In this case, $\max_{j} \mathcal{A}_{ij}$, which would be the peak-to-peak amplitude that is compared to the threshold, comes from this bad sensor. If the sensor is not repaired or removed, we might end up rejecting a large fraction of otherwise good trials, just because of a single bad sensor. This is certainly not optimal. In fact, a possibly better solution is to replace the corrupted signal in the sensor by the interpolation of the signals in the nearby sensors. A second observation is that sensors can have very different ranges of amplitudes depending on their location on the scalp. A threshold tuned for one sensor may not work as effectively for another sensor. Both of these observations are motivations for estimating rejection thresholds for each sensor separately.

Once we define sensor-wise rejection thresholds $\tau_{\star}^{j}$, we can define an indicator matrix $C_{ij} \in \{0, 1\}^{N \times Q}$ which designates the bad trials at the level of individual sensors. In other words, we have:
\begin{equation}
C_{ij} = \begin{cases} 
0, & \text{if } \mathcal{A}_{ij} \leq \tau^{j}_{\star} \\
1, & \text{if } \mathcal{A}_{ij} > \tau^{j}_{\star}
\end{cases}
\end{equation}
The schematic in Figure~\ref{fig:schematic}A shows a cartoon figure for this indicator matrix $C_{ij}$. Now that we have identified bad sensors for each trial, one might be tempted to interpolate all the bad sensors in each trial. However, it is not as straightforward since in some trials, a majority of the sensors may be bad. These trials cannot be repaired by interpolation and must be rejected. In some other cases, the number of bad sensors may not be large enough to justify rejecting the trial. However, it might already be too much to interpolate all the sensors reliably. In these cases, a natural idea is to pick the worst few sensors and interpolate them. This suggests an algorithm as described in Figure~\ref{fig:schematic}B. Reject a trial only if most sensors ``agree" that the trial is bad, otherwise interpolate as many sensors as possible. We will denote by $\kappa$ the maximum number of bad sensors in a non-rejected trial and by $\rho$ the maximum number of sensors that can be interpolated. Note that $\rho$ is necessarily less than $\kappa$. The interpolation scheme for EEG uses spherical splines~\citep{perrin1989spherical} while for MEG it uses a Minimum Norm Estimates formulation with spherical harmonics~\citep{hamalainen1994interpreting}. The implementation is provided by MNE-Python~\citep{gramfort2013meg}.

The set of good trials $\mathcal{G}^{\kappa}_k$ in the training set $\mathcal{T}_k$ can therefore be written mathematically as:
\begin{equation}
\mathcal{G}^{\kappa}_{k} = \{i \in \mathcal{T}_k \suchthat \sum_{j=1}^{Q} C_{ij} < \kappa \} \enspace .
\end{equation}
In the remaining trials, if $\rho < \kappa$, one needs to define what are the worse $\rho$ sensors that shall be interpolated. To do this we propose to rank the sensors for ``badness'' according to a score. A natural strategy to set the score is to use the peak-to-peak amplitude itself:
\begin{equation}
s_{ij} = \begin{cases}
\mathcal{A}_{ij} & \text{if } C_{ij} = 1 \\
-\infty & \text{if } C_{ij} = 0
\end{cases}
\label{eq:score}
\end{equation}

Higher the score $s_{ij}$, the worse is the sensor. The $-\infty$ score is for ignoring the good sensors in the subsequent step. The following strategy is used for interpolation.
If the number of bad sensors $\sum_{j'=1}^{Q} C_{ij'}$ is less than $\rho$ we will interpolate all of them. Otherwise, we will interpolate the $\rho$ sensors with the highest scores.
In other words, we interpolate at most $\mathrm{min}(\rho, \sum_{j'=1}^{Q} C_{ij'})$ sensors.

Denoting by $X^{\rho}_{\mathcal{G}^{\kappa}_k}$ the data in the training set after rejection and cleaning by interpolation, the RMSE averaged over $K$ folds for the parameter pair $(\rho, \kappa)$ therefore becomes:
\begin{equation}
\overbar{e}(\rho, \kappa) = \frac{1}{K} \sum_{k=1}^{K} \fro{\overbar{X^{\rho}_{\mathcal{G}^{\kappa}_k}} - \widetilde{X_{\mathcal{V}_k}}}
\end{equation}
where $\fro{\cdot}$ is the Frobenius norm.
Finally, the best parameters $\rho_{*}$ and $\kappa_{*}$ are estimated using grid search~\citep{hsu2003practical}.

\subsubsection{Data augmentation}
\label{sec:data_augmentation}

\begin{figure}[ht!]
    \centering
    \includegraphics[width=0.7\linewidth]{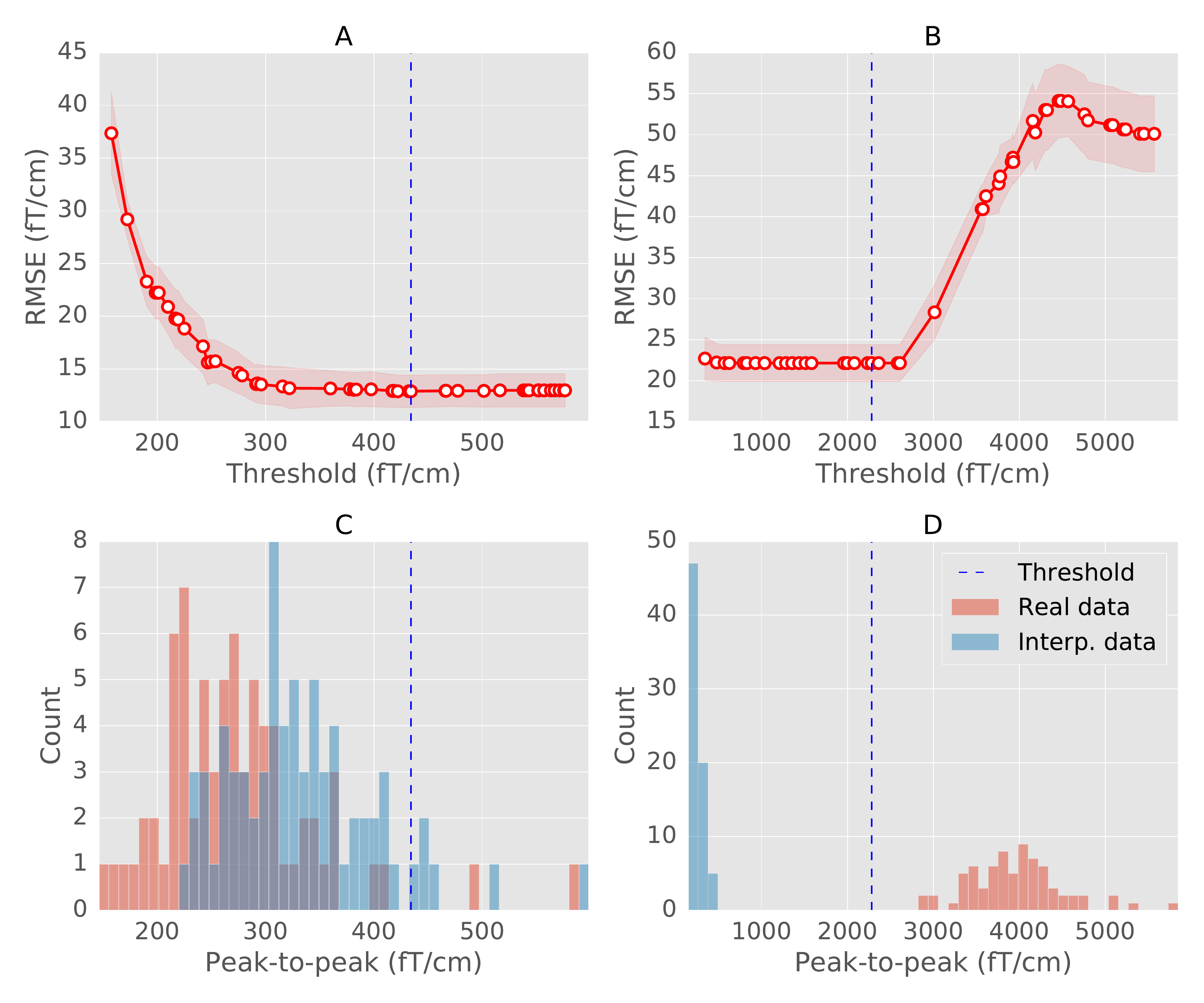}
    \caption{(A) and (B) The cross-validation curve obtained with sequential Bayesian optimization (see Section~\ref{sec:bayesian_opt} for an explanation) for a regular (MEG 2523) and a globally bad sensor (MEG 2443) from the MNE sample dataset. The mean RMSE is shown in red circles with error bounds in red shades. The red shaded region shows the lower and upper bounds between which the optimization is carried out. Vertical dashed line marks the estimated threshold. (C) and (D) Histogram of peak-to-peak amplitudes of trials in the sensor. The histograms are computed separately for the real data (red) and the data interpolated from other sensors (blue). The estimated threshold correctly marks all the trials as bad for the globally bad sensor.}
    \label{fig:cross_val_hist}
\end{figure}

In practice, cross-validation does not work for a globally bad sensor since all the trials are corrupted. In this scenario, the optimal threshold for this bad sensor should be lower than the lowest peak-to-peak amplitude so that all the trials for that sensor are marked as bad. However, even the median of the validation set has been corrupted. The algorithm therefore attempts to keep as many trials as necessary for the average to be close to the corrupted median. Thus, the estimated threshold ends up being higher than what would have been optimal. Recall from Figure~\ref{fig:cross_val} that this is the classic case of an overfitting model. A common strategy in machine learning to reduce overfitting is data augmentation~\citep{krizhevsky2012imagenet}. It basically boils down to using the properties of the data, such as the physics of the system, to generate more plausible data.

To implement data augmentation in our model, we interpolate each sensor from all the other $Q-1$ sensors and by doing so, we double the number of trials in the data. In the augmented data, half of the trials contain sensor data which are the output of a leave-one-sensor-out cross-validation. The augmented data matrix is $X^{\textrm{aug}} \in \real^{2N \times P}$. With the augmented data, the median is now closer to the uncorrupted median of the data in that sensor. During cross-validation the folds were stratified so that the number of interpolated trials and original trials in each fold were roughly equal.

\subsection{Candidate thresholds using Bayesian optimization}
\label{sec:bayesian_opt}
Now that we have formalized the problem and our approach, we must estimate the threshold $\tau_{\star}$ which minimizes the error defined in Equation~\eqref{eq:err}. A na\"ive strategy is to define a set of equally spaced points over a range of thresholds $[\tau_{\min}, \tau_{\max}]$. The estimated threshold would be the one which obtains the lowest error among these candidate threshold. This is the approach taken in Figure~\ref{fig:cross_val}. The range of thresholds is easy to set as it can be determined from the minimum and maximum peak-to-peak amplitude for the sensor in the augmented data matrix $X^{\textrm{aug}}$. However, it is not obvious how to set the spacing between the candidate thresholds, and experiments showed that varying this spacing could impact the results. If the candidate thresholds are far apart, one might end up missing the optimal threshold. On the other hand, if the thresholds are very dense, it is computationally more demanding.

This motivated us to use Bayesian optimization~\citep{snoek2012practical, bergstra2011algorithms} to estimate the optimal thresholds. It is a sequential approach which decides the next candidate threshold to try based on all the observed thresholds so far. It is based on maximizing an acquisition function given an objective function of samples seen so far (data likelihood) and the prior (typically a Gaussian Process (GP)~\citep{rasmussen2006gaussian}). The objective function in our case is the mean cross-validation error as defined in Equations~\eqref{eq:err}. To obtain the next iterate, an acquisition function is maximized over the posterior distribution. Popular choices of the acquisition function include ``probability of improvement", ``expected improvement" and ``confidence bounds of the GP"~\citep{snoek2012practical}. We pick ``expected improvement" as it balances exploration (searching unknown regions) and exploitation (maximizing the improvement) strategies without the need of a tuning parameter. For our analysis, we use the scikit-optimize\footnote{https://scikit-optimize.github.io} implementation of Bayesian optimization, which internally uses the Gaussian process module from scikit-learn~\citep{scikit-learn}.

Figure~\ref{fig:cross_val_hist}A and \ref{fig:cross_val_hist}B show the cross-validation curve for a regular sensor and a globally bad sensor in the MNE sample dataset~\citep{mne,gramfort2013meg}. The RMSE is evaluated on thresholds as determined by the Bayesian optimization rather than a uniform grid. These plots also illustrate the arguments presented in Section~\ref{sec:data_augmentation} with respect to data augmentation. The histograms in Figure~\ref{fig:cross_val_hist}C for the interpolated data and the real data are overlapping for the regular sensor. Thus, the estimated threshold for that sensor marks a trial as outlier if its peak-to-peak values is much higher than the rest of the trials. However, in the case of a globally bad sensor, the histogram (Figure~\ref{fig:cross_val_hist}D) is bimodal -- one mode for the interpolated data and one mode for the real data. Now, the estimated threshold is no longer marking outliers in the traditional sense. Instead, all the trials belonging to that sensor must be marked as bad.

%% file: sections/validation_protocol.tex
\section{Experimental Validation Protocol}

\begin{table}[!b]
{
    \caption{Overview of rejection strategies evaluated\label{tab:strategies}}
       \begin{center}
       \begin{tabular}{l l l l}
        \hline
        \textbf{method} & \textbf{statistical scope} & \textbf{parameter defaults}\\
        \hline
        $\text{FASTER}^{a}$ & univariate & threshold on zscore $=$ 3 \\
        $\text{SNS}^{b}$ & multivariate & number of neighbors = 8\\
        $\text{RANSAC}^{c}$ & multivariate outlier detection & \#resamples = 50, fraction of channels = 0.25,\\
        & & threshold on correlation = 0.75, unbroken time = 0.4 \\
        autoreject & univariate with cross-validation & sensor-level thresholds, $\rho$ and $\kappa$; learned from data \\
        \hline
        \end{tabular}
        \label{table:methods}
        \end{center}
        \vspace{-0.9em}
        \hspace{1em}
        {\footnotesize
         $^a$\cite{nolan2010faster}, $^b$\cite{de2008sensor},  $^c$\cite{bigdely2015prep}}
}
\end{table}

To experimentally validate \emph{autoreject}, our general strategy is to first visually evaluate the results and thereafter quantify the performance. We describe below the evaluation metric used, the methods we compare against, and finally the datasets analyzed. All general data processing was done using the open source software MNE-Python~\citep{gramfort2013meg}.

\subsection{Evaluation metric}
The evoked response from the data cleaned using our algorithm or a competing benchmark is denoted by $\overbar{X}(method)$. This is compared to the ground truth evoked response $\overbar{X}(clean)$ (See Section~\ref{sec:datasets} to see how these are obtained for different datasets) using:
\begin{equation}
\infnorm{\overbar{X}(method) - \overbar{X}(clean)}
\label{eq:infnorm}
\end{equation}
where $\infnorm{\cdot}$ is the infinity norm. The reason for using infinity norm is that it is sensitive to the maximum amplitude in the difference signal as opposed to the Frobenius norm which averages the squared difference. The $\infnorm{\cdot}$ is a particularly sensitive metric to quantity artifacts which are also visually striking such as those localized on one sensor or at a given time instant.

\subsection{Competing methods}
\label{sec:competing_methods}

Here, we list the methods that will be quantitatively compared to \emph{autoreject} using the evaluation metric in Equation~\ref{eq:infnorm}. These methods are also summarized for the reader's convenience in Table~\ref{table:methods}.

\begin{itemize}[noitemsep,nolistsep]
\item \emph{No rejection}: It is a simple sanity check to make sure that the data quality upon applying the \emph{autoreject (local)} algorithm does indeed improve. This is the data before the algorithm is applied.
\item \emph{Sensor Noise Suppression (SNS)}: The SNS~\citep{de2008sensor} algorithm, as described in the Introduction (Section~\ref{sec:introduction}), projects the data of each sensor on to the subspace spanned by the principle components of all the other sensors. What it does is regressing out the sensor noise that cannot be explained by other sensors. It works on the principle that brain sources project on to multiple sensors but the noise is uncorrelated across sensors. In practice, not all the sensors are used for projection, but only a certain number of neighboring sensors (determined by the correlation in the data between the sensors).
\item \emph{Fully Automated Statistical Thresholding for EEG artifact Rejection (FASTER)}: It finds the outlier sensor using five different criteria: the variance, correlation, Hurst exponent, kurtosis and line noise. When the z-score of any of these criteria exceeds 3, the sensor is marked as bad according to that criteria. Note that even though FASTER is typically used as an integrated pipeline, here we use the bad sensor detection step, as this is what appears to dominate the bad signals in the case of the HCP data (Section~\ref{sec:datasets}). We take a union of the sensors marked as bad by the different criteria and interpolate the data for those sensors.
\item \emph{Random Sample Consensus (RANSAC)}: We use the RANSAC implemented as part of the PREP pipeline~\citep{bigdely2015prep}. In fact, RANSAC~\citep{fischler1981random} is a well-known approach used to fit statistical models in the presence of outliers in the data. In this approach, adopted for the use case of artifact detection in EEG, a subset of sensors (inliers) are sampled randomly (25\% of the total sensors) and the data in all sensors are interpolated from these inliers sensors. This is repeated multiple times (50 in the PREP implementation) so as to yield a set of 50 time series for each sensor. The correlation between the median, computed instant by instant, of these 50 time series and the real data is computed. If this correlation is less than a threshold (0.75 in the PREP implementation), then the sensor is considered an outlier and therefore marked as bad. It is perhaps worth noting that unlike in the classical RANSAC algorithm, the inlier model is not learned from the data but instead determined from the physical interpolation. A sensor which is bad for more than 40\% of the trials (the unbroken time) is marked as globally bad and interpolated. Even though the method was first proposed on EEG data only, we extended it for MEG data by replacing spline interpolation with field interpolation using spherical harmonics as implemented in MNE~\citep{gramfort2013meg,hamalainen1994interpreting}. Note that this is the same interpolation method that is used by \emph{autoreject (local)}.
\end{itemize}

\subsubsection{Datasets}
\label{sec:datasets}

\begin{table}[!t]
{
    \caption{Overview of datasets analyzed\label{tab:datasets}}
    \resizebox{\textwidth}{!}{
        \begin{tabular}{l l l l l l}
        \hline
         \textbf{Algorithm} & \textbf{Dataset} & \textbf{Acquisition device} & \textbf{Sensors used} & \textbf{\#subjects}\\

\hline
\multirow{2}{*}{autoreject (global)} & MNE sample data & Neuromag VectorView & 60 EEG electrodes & 1\\
& EEGBCI & BCI2000 cap & 64 EEG electrodes & 105\\
\hline
\multirow{3}{*}{autoreject (local)} & MNE sample data & Neuromag VectorView & 60 EEG electrodes & 1\\
& EEG faces & Neuromag VectorView& 60 EEG electrodes & 19\\
& HCP working memory & 4D Magnes 3600 WH& 248 magnetometers & 83\\
        \hline
        \end{tabular}
    }
    \label{table:datasets}
}
\end{table}

We validated our methods on four open datasets with data from over 200 subjects. This allowed us to evaluate experimentally strengths and potential limitations of different rejection methods. The datasets contained either EEG or MEG data. To obtain solid experimental conclusions, diverse experimental paradigms were considered with data from working memory, perceptual and motor tasks.

We detail below how we defined $\overbar{X}(clean)$, the cleaned ground-truth data for two of our datasets -- HCP MEG and EEG faces data. This is perhaps one of the most challenging aspects of this work because the performance is evaluated on real data and not on simulations. An overview of all the datasets used in this study is provided in Table~\ref{table:datasets}.

\paragraph{MNE sample data}

The MNE sample data~\citep{gramfort2013meg} is a multimodal open dataset consisting of MEG and EEG data. It has been integrated as the default testing dataset into the development of the MNE software~\citep{gramfort2013meg}. The simultaneous M/EEG data were recorded at the Martinos Center of Massachusetts General Hospital. The MEG data with a Neuromag VectorView system, and an MEG-compatible cap comprising 60 electrodes was used for the EEG recordings. Data were sampled at 150 Hz. In the experiment, auditory stimuli (delivered monoaurally to the left or right ear) and visual stimuli (shown in the left or right visual hemifield) were presented in a random sequence with a stimulus onset asynchrony of 750 ms. The data was low-pass filtered at 40 Hz. The trials were 700 ms long including a 200 ms baseline period which was used for baseline correction.

\paragraph{EEGBCI dataset}

This is a 109-subject dataset (of which we analyzed 105 subjects which can be easily downloaded and analyzed using MNE-Python~\citep{gramfort2013meg}) containing EEG data recording with a 64-sensor BCI2000 EEG cap~\citep{schalk2004bci2000}. Subjects were asked to perform different motor/imagery tasks while their EEG activity was recorded. In the related BCI protocol, each subject performed 14 runs, amounting to a total of 180 trials for hands and feet movements (90 trials each). The data was band-pass filtered between 1 and 40 Hz, and 700 ms long trials were constructed including a 200 ms pre-stimulus baseline period.

\paragraph{EEG faces data (OpenfMRI ds000117)}

The OpenfMRI ds000117 dataset~\citep{wakeman2015multi} contains multimodal task-related neuroimaging data over multiple runs for EEG, MEG and fMRI. For our analysis, we restrict ourselves to EEG data. The EEG data was recorded using a 70 channel Easycap EEG with electrode layout conforming to the 10-10\% system. Subjects were presented with images of famous faces, unfamiliar faces and scrambled faces as stimuli. For each subject, on average, about 293 trials were available for famous and unfamiliar faces. The authors kindly provided us with run-wise bad sensor annotations which allowed us to conduct benchmarking against human judgement. To generate the ground truth evoked response $\overbar{X}(clean)$, we randomly select 80 percent of the total number of trials in which famous and unfamiliar faces were displayed. In these trials, we interpolated the bad sensors run-wise. Then, we removed physiological artifacts (heart beat and eye blinks) using Independent Component Analysis (ICA)~\citep{vigario2000independent}. Following the ICA pipelines recommended by the MNE-Python software, the bad ICA components were marked automatically using cross-trial phase statistics~\citep{dammers2008integration} for ECG (threshold=0.8) and adaptive z-scoring (threshold=3) for EOG components. The evoked response from the cleaned data $\overbar{X}(method)$ is computed from the remaining 20 percent trials cleaned using either \emph{autoreject (local)} or \emph{RANSAC} (see Section~\ref{sec:benchmark_sensors} for a description of this method). Computing the ground-truth evoked potential from a large proportion of trials minimized the effect of outliers in the average. However, it is noteworthy that this choice of assigning fewer trials to the estimation with rejection algorithms acts in a conservative sense: each unnoticed bad trial may affect the ensuing evoked potentials more severely.

\paragraph{Human Connectome Project (HCP) MEG data}

The HCP dataset is a multimodal reference dataset realized by the efforts of multiple international laboratories around the world. It currently provides access to both task-free and task-related data for more than 900 human subjects with functional MRI data, 95 of which have presently also MEG~\citep{larson2013adding}. An interesting aspect of the initiative is that the data provided is not only in unprocessed BTi format, but also processed using diverse processing pipelines. These include annotations of bad sensors and corrupted time segments for the MEG data derived from automated pipelines and supplemented by human inspection. The automated pipelines are based on correlation between neighboring sensors, z-score metrics, ratio of variance to neighbors, and ICA decomposition. Most significant for our purposes, the clean average response $\overbar{X}(clean)$ is directly available. It allows us to objectively evaluate the proposed algorithm against state-of-the-art methods by reprocessing the raw data and comparing the outcome with the official pipeline output.

The HCP MEG dataset provides access to MEG recordings from diverse tasks, \textit{i.e.}, a motor paradigm, passive listening and working memory. Here, we focused on the working memory task for which data is available for 83 subjects out of 95. A considerable proportion of subjects were genetically related, but we can ignore this information as the purpose of our algorithm is artifact removal rather than analyzing brain responses. For each subject two runs are available. Two classes of stimuli were employed, faces and tools. Here, we focused on the MEG data in response to stimulus onsets for the ``faces" condition.

The MEG data were recorded with a wholehead MAGNES 3600 (4D Neuroimaging, San Diego, CA) in a magnetically shielded room at Saint Louis University. The system comprises 248 magnetometers and 23 reference sensors to capture environmental signals. Time windows precisely matched values used by the HCP ``eravg'' pipeline with onsets and offsets at $-1.5$\,s and $2.5$\,s before and after the stimulus event, respectively. As in the HCP pipeline, signals were down-sampled to $508.63$\,Hz and band-pass filtered between 0.5--60\,Hz. As it is commonly done with BTi systems, reference sensors at the periphery of the head were used to subtract away environmental noise. Given the linearity of Maxwell equations in the quasi-static regime, a linear regression model was employed. More precisely, signals from reference sensors are used as regressors in order to predict the MEG data of interest. The ensuing signal explained by the reference sensors in this model was then removed. The HCP preprocessing pipeline contains two additional steps: ICA was used to remove components not related to brain activity (including eye blinks and heart beats) and then bad trials and bad segments were removed with a combination of automated methods as well as annotations by a human observer. To have a fair comparison and focus on the latter step, the ICA matrices provided by the HCP consortium were applied to the data. We interpolated the missing sensors in $\overbar{X}(clean)$ so that it has the same dimensions as the data from $\overbar{X}(method)$. All the algorithms were executed separately on each run and the evoked response of the two runs was averaged to get $\overbar{X}(method)$.

To enable easy access of the files along with compatibility in MNE-Python, we make use of the open source MNE-HCP package\footnote{http://mne-tools.github.io/mne-hcp/}. For further details on the HCP pipelines, the interested reader can consult the related paper by \citet{larson2013adding} and the HCP S900 reference manual for the MEG3 release.

%% file: sections/results.tex
\section{Results}
\label{sec:results}

We conducted qualitative and quantitative performance evaluations of \emph{autoreject} using four different datasets comparing it to a baseline condition without rejection as well as three different alternative artifact rejection procedures.

\subsection{Peak-to-peak thresholds}

\begin{figure}[htb!]
	\centering
	\includegraphics[width=0.9\linewidth]{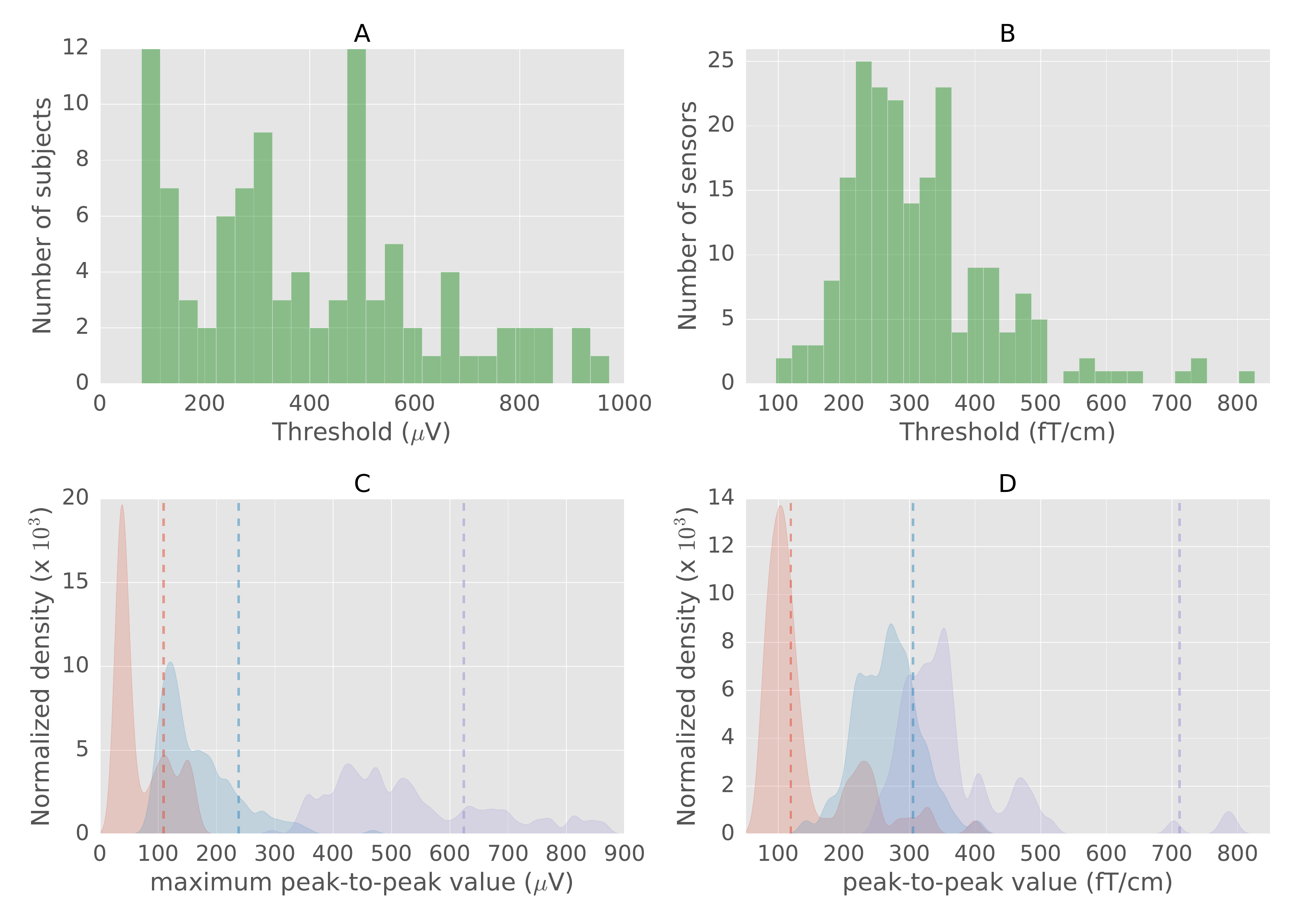}
    \caption{A. Histogram of thresholds for subjects in the EEGBCI dataset with \emph{autoreject (global)} B. Histogram of sensor-specific thresholds in gradiometers for the MNE sample dataset (Section~\ref{sec:results}). C. Normalized kernel density plots of maximum peak-to-peak value across sensors for three subjects in the EEGBCI data. Vertical dashed lines indicate estimated thresholds. Density plots and thresholds corresponding to the same subject are the same color. D. Normalized Kernel Density plots of peak-to-peak values for three MEG sensors in the MNE sample dataset. The threshold indeed has to be different depending on the data (subject and sensor).}
    \label{fig:hist}
\end{figure}

First, let us convince ourselves that the peak-to-peak thresholds indeed need to be learned. In Figure~\ref{fig:hist}A, we show a histogram of the thresholds learned on subjects in the EEGBCI dataset using \emph{autoreject (global)}. This figure shows that thresholds vary a lot across subjects. One could argue that this is due to variance in the estimation process. To rule out such a possibility, we plotted the distribution of maximum peak-to-peak thresholds as kernel density plots in Figure~\ref{fig:hist}C for three different subjects. We can see that these distributions are indeed subject dependent, which is why a different threshold must be learned for each subject. In fact, if we were to use a constant threshold of $150 \mu{V}$, in $17\%$ of the subjects, all the trials would be dropped in one of the two conditions. Of course, from Figure~\ref{fig:hist}A, we can now observe that $150 \mu{V}$ is not really a good threshold to choose for many subjects.

We show here the maximum peak-to-peak amplitude per sensor because this is what decides if a trials should be dropped or not in the case of \emph{autoreject (global)}. Note that, if instead, we examined the distribution of peak-to-peak amplitudes across all sensors and trials, we would see a quasi-normal distribution. When all the sensors are taken together, a ``smoothing" effect is observed in the distribution. This is a consequence of the central limit theorem. This also explains why we cannot learn a global threshold using all the peak-to-peak amplitudes across trials and sensors.

With the \emph{autoreject (local)} approach, a threshold is estimated for each sensor separately. The histogram of thresholds for the MNE sample dataset is plotted in Figure~\ref{fig:hist}B. It shows that the threshold varies even across homogeneous MEG sensors. Figure~\ref{fig:hist}D shows the distribution of peak-to-peak thresholds for three different MEG sensors. This graph confirms actual sensor-level differences in amplitude distributions, which was also previously reported in the literature~\citep{junghofer2000statistical}. With this work, we go one step further by learning automatically the thresholds in a data-driven way rather than asking users to mark them interactively.

\begin{figure}[htb!]
	\centering
	\includegraphics[width=0.8\linewidth]{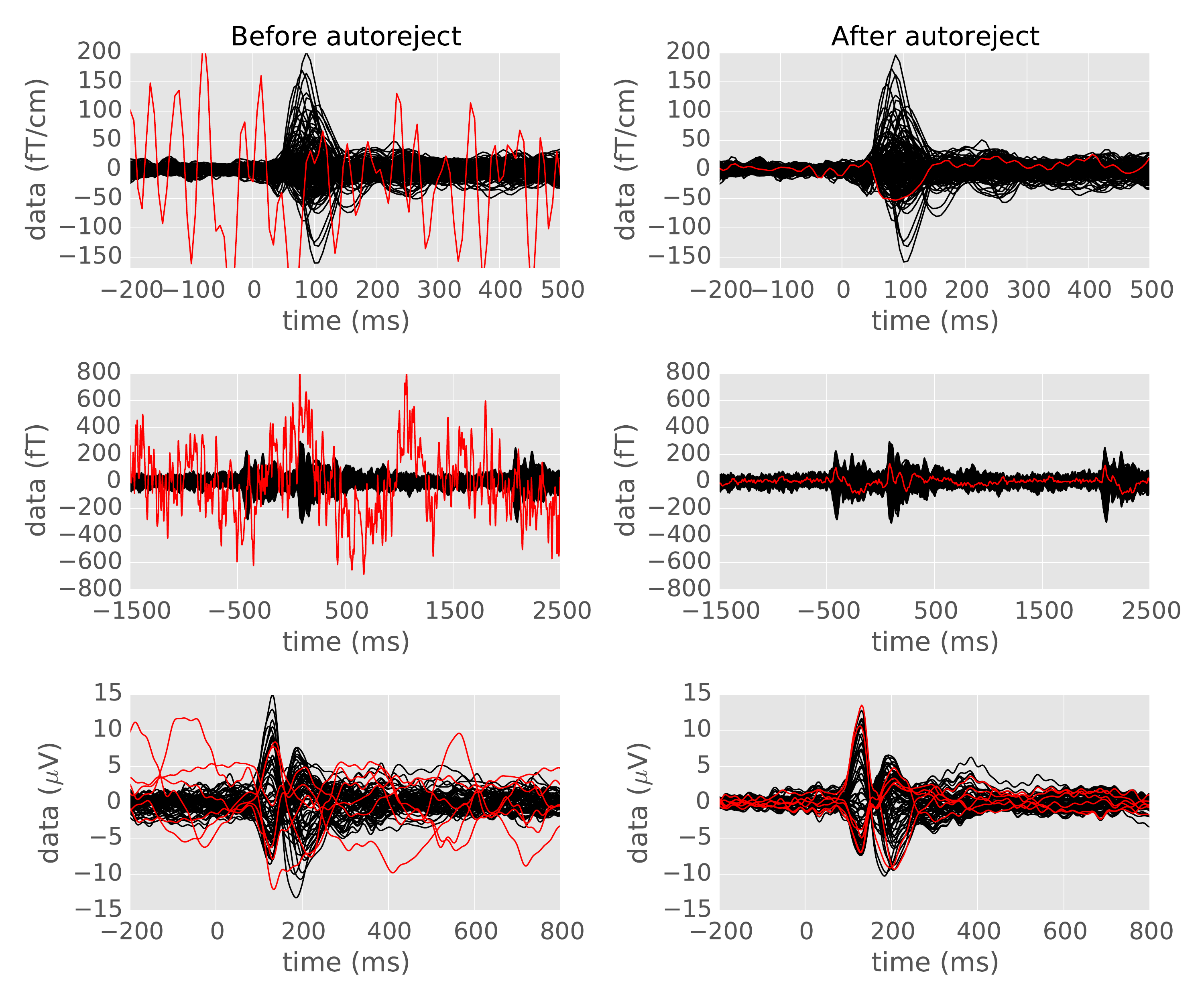}
    \caption{The evoked response (average of data across trials) on three different datasets before and after applying \emph{autoreject} --- the MNE sample data, the HCP data and the EEG faces data. Each sensor is a line on the plots. On the left, manually annotated bad sensors are shown in red. The algorithm finds the bad sensors automatically and repairs them for the relevant trials. Note that it can even fix multiple sensors at a time and works for different modalities of data acquisition.}
    \label{fig:sample_evoked}
\end{figure}

\subsection{Visual quality check}

\begin{figure}[htb!]
    \centering
    \includegraphics[width=0.75\linewidth]{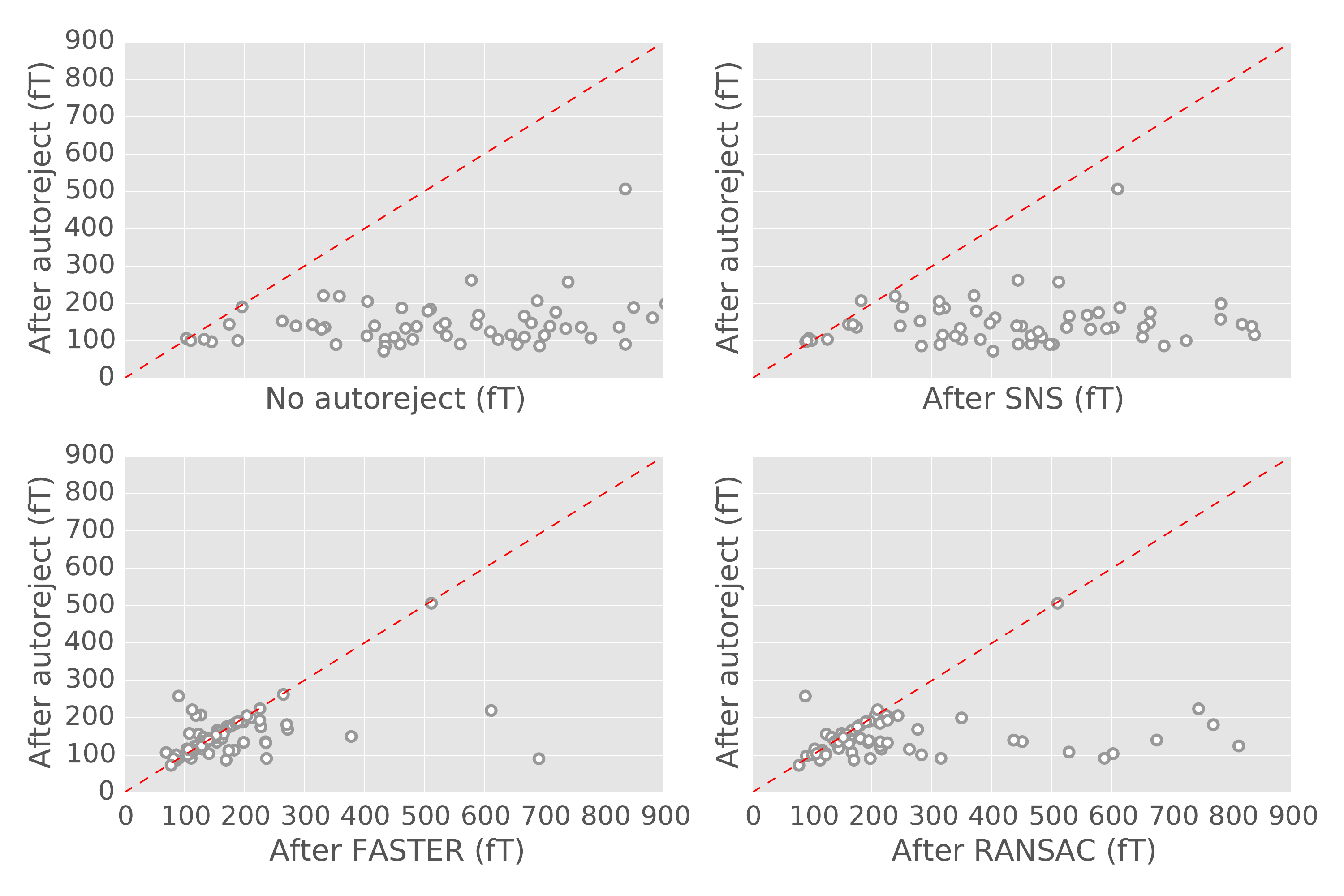}
    \caption{Scatter plots for the results with the HCP data. For each method, the $\infnorm{\cdot}$ norm of the difference between the HCP ground truth and the method is taken. Each circle is a subject. (A) \textit{autoreject (local)} against no rejection, (B) \textit{autoreject (local)} against Sensor Noise Suppression (SNS) (SNS), (C) \textit{autoreject} against FASTER, (D) \textit{autoreject (local)} against RANSAC. Data points below the dotted red line indicate subjects for which \textit{autoreject (local)} outperforms the alternative method.}
    \label{fig:hcp_scatter}
\end{figure}

\begin{figure}[htb!]
    \centering
    \includegraphics[width=0.75\linewidth]{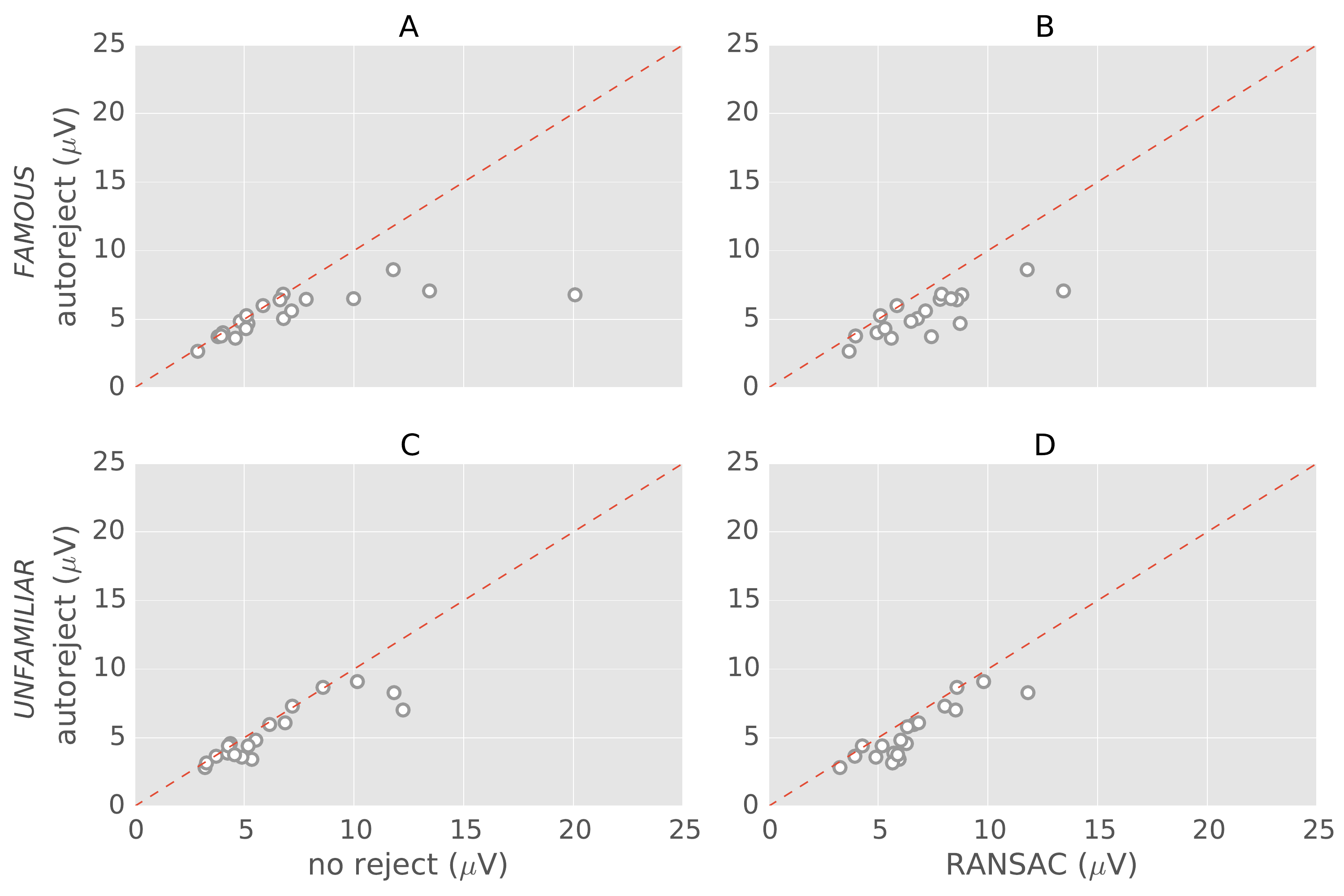}
    \caption{Scatter plots for the results with the 19 subjects from Faces dataset. The first row (A) and (B) is for the condition ``famous" and the second row (C) and (D) is for the condition ``unfamiliar" faces. For each method, the $\infnorm{\cdot}$ norm of the difference between the ground truth and the estimates is computed. Each circle is a subject. Data points below the dotted red line indicate subjects for which \textit{autoreject (local)} outperforms the alternative method.}
    \label{fig:dgw_scatter}
\end{figure}

The average response plotted in a single graph, better known as ``butterfly plots'', constitutes a natural way to visually assess the performance of the algorithm for three different datasets -- MNE sample data, HCP MEG data, and EEG faces data. In Figure~\ref{fig:sample_evoked}, the subplots in the left column show the evoked response with the bad sensors marked in red. Right subplots, show data after applying the \emph{autoreject (local)} algorithm, with the repaired bad sensors in red. The algorithm works for different acquisition modalities -- MEG and EEG, and even when multiple sensors are bad. A careful look at the results, show that \emph{autoreject (local)} does not completely remove eyeblinks in the data as some of the blinks are time-locked to the evoked response. We will later discuss (Section~\ref{sec:discussion}) the possible solutions of applying ICA-based artifact correction in combination with \emph{autoreject (local)}.

\subsection{Quantification of performance and comparison with state-of-the-art}
\label{sec:benchmark_sensors}

We now compare these algorithms to \emph{autoreject (local)} using the data quality metric defined in Equation~\eqref{eq:infnorm}. We are interested not only in how the algorithms perform on average but at the level of individual subjects. To detail single subject performance, we present the data quality as scatter plots where each axis corresponds to the performance of a method. Figure~\ref{fig:hcp_scatter}, contains results on the HCP MEG data. We can observe from the top-left subplot of the figure that \emph{autoreject (local)} does indeed improve the data quality in comparison to the \emph{no rejection} approach. In Figure~\ref{fig:hcp_scatter}B, \emph{autoreject (local)} is compared against SNS. The SNS algorithm focuses on removing noise isolated on single sensors. Its results can be affected by the presence of multiple bad sensors and globally bad trials. This explains why \emph{autoreject (local)} outperforms SNS is this setting. In Figure~\ref{fig:hcp_scatter}C, we compare against FASTER. Even though \emph{autoreject (local)} is slightly worse than FASTER for a few subjects, FASTER is clearly much worse than \emph{autoreject (local)} for at least 3 subjects, and \emph{autoreject (local)} yields therefore less errors on average. Finally, Figure~\ref{fig:hcp_scatter}D shows comparison to RANSAC. In the PREP implementation, this algorithm is not fully data-driven in the classic sense of RANSAC. This is due to the fact that the inlier model is not learned but rather derived from the physics of the interpolation. It is therefore an algorithm which is conceptually close to \emph{autoreject}. However, a critical difference is that the parameters of this method still need to be tuned. This can be a problem as these parameters can be suboptimal on some datasets. Some experiments showed that it is for example the case for the EEG faces data, where it is possible to obtain better results by manually tuning the RANSAC parameters, rather than using the values proposed by the original authors.

Figure~\ref{fig:dgw_scatter} presents scatter plots for the EEG faces data.
Here, we restrict our comparison to RANSAC as it is conceptually the closest to \emph{autoreject}. On this data, we apply the algorithms on both the conditions -- famous and unfamiliar faces. It should be noted that the ground truth for this data was generated automatically with no additional annotations from human experts. However, a sanity check was performed on the ground truth by visual inspection. Here too, \emph{autoreject} offers good results across all subjects, and even for the subjects for which RANSAC underperforms.

\clearpage
\begin{sidewaysfigure}
	\centering
	\includegraphics[width=\textwidth]{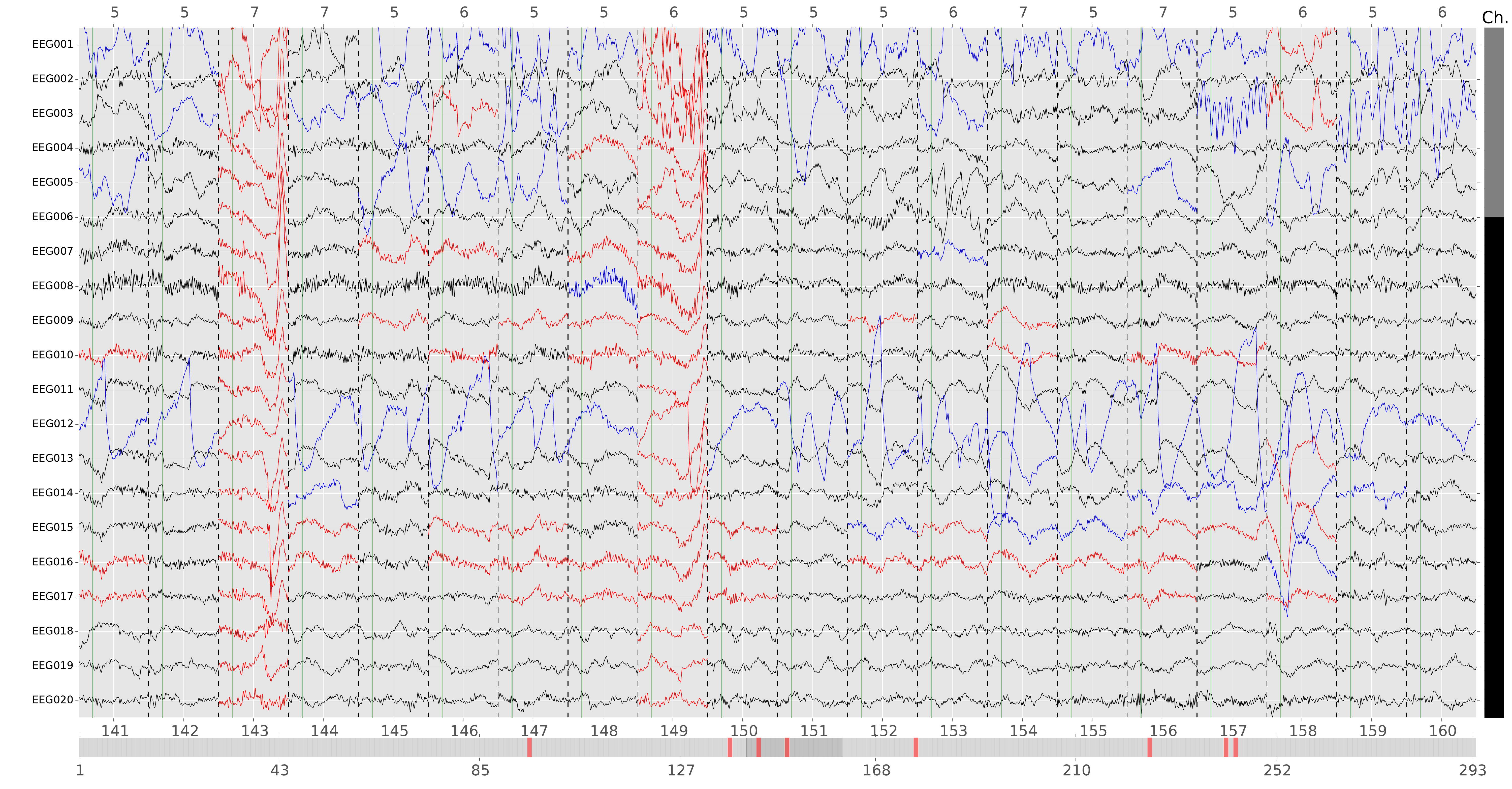}
    \caption{An example diagnostic plot from an interactive viewer with \emph{autoreject (local)}. The data plotted here is subject 16 for the condition `famous' in the EEG faces data. Each row is a different sensor. The trials are concatenated along the x axis with dotted vertical lines separating consecutive trials. Each trial is numbered at the bottom and its corresponding trigger code is at the top. The horizontal scroll bar at the bottom allows browsing trials and the vertical scroll bar on the right is for browsing sensors. A trial which is marked as bad is shown in red on the horizontal scroll bar and the corresponding column for the trial is also red. A data segment in a good trial is either i) Good (in black) ii) Bad and interpolated (blue), or iii) Bad but not interpolated (in red). Note that the worst sensors in a trial are typically interpolated.}
    \label{fig:diagnostic_plot}
\end{sidewaysfigure}
\clearpage

%% file: sections/discussion.tex
\section{Discussion}
\label{sec:discussion}

In this study, we have presented a novel artifact rejection algorithm called \emph{autoreject} and assessed its performance on multiple datasets showing comparisons with other methods from the state-of-the-art.

We have shown that learning peak-to-peak rejection thresholds subject-wise is justified as the distribution of this statistic indeed varies considerably across subjects. We have shown qualitatively that \emph{autoreject} yielded clean physiological event related field (ERF) and event related potentials (ERP) by correcting or rejecting contaminated data segments. Finally, we have shown quantitavely that \emph{autoreject} yields results closer to the ground truth for more subjects than the algorithms presented in Section~\ref{sec:competing_methods}. We now further discuss the conceptual similarities and differences of our approach to the alternative methods. We also discuss the interaction between \emph{autoreject}
and some other steps in the M/EEG analysis pipelines.

\subsection{Autoreject vs. competing methods}

We believe the key advantage of \emph{autoreject (local)} over the other methods consists in combining data-driven parameter tuning with deterministic and physics-driven data interpolation. This interpolation promotes spatial smoothness of the electric potential on the scalp for EEG, and in the case of MEG, explicitly takes into account the well-understood Maxwell's equations. To recapitulate, the sensor-level thresholds mark outlier segments across trials at the level of individual sensors, following a data augmentation step which exploits the full array of sensors. As trials are seen as independent observations, the thresholds can be therefore learned from the data using cross-validation. The cross-validation is stratified so that each fold contains roughly an equal proportion of the original and augmented trials. At repair time, bad segments are replaced with interpolated data from the good sensors. Of course, this is problematic if the sensor locations are not readily available. Fortunately, it turns out that the sensor positions from standard montages are often good enough for reliable interpolation.

In contrast to \emph{autoreject (local)}, SNS is a purely statistical method that does not take into account the physics of sensor locations for repairing the data. In SNS, the sensors are considered in a leave-one-sensor-out protocol. For each sensor, a ``clean'' subspace is defined from the principle components of the remaining sensors. The data from this sensor is then projected on to the ``clean'' subspace. As we have seen in Section~\ref{sec:results} (Figure~\ref{fig:hcp_scatter}), this does not work satisfactorily, presumably because the SNS method makes strong assumptions regarding the orthogonality of the noise and ``clean'' subspace. The ensuing projection may not improve, and even deteriorate the signal in some cases. The consequence of this is what we observe empirically in Figure~\ref{fig:hcp_scatter}. Applying SNS will also be problematic when multiple sensors are corrupted simultaneously. However, this is less of a problem in the HCP MEG data that we analyzed. 

On the other hand, the FASTER method derives its rejection decisions from multiple signal characteristics. It uses criteria such as between-sensor correlation, variance and power spectrum, by considering their univariate Gaussian statistics with thresholds fixed to a z-score of 3. This default threshold appears to be satisfying as they work on a vast majority of subjects. However, the fact that it does not work as well on certain subjects can limit its adoption for large scale studies. Here, the adaptive nature of threshold detection performed by \emph{autoreject} seems to be a clear advantage.

The RANSAC algorithm also performs adaptive outlier detection, but across sensors rather than trials. While \emph{autoreject (local)} operates on segmented data such as trials time-locked to the stimuli, RANSAC was designed for continuous data without any segmentation. In fact, one could readily obtain bad sensor per trial (as illustrated in Figure~\ref{fig:schematic}) even with RANSAC. However, the authors of the paper did not validate their method on continuous data, and hence, such a modification would require additional work. Although in the case of MEG data, this is not very crucial, this can in fact be critical for EEG data analysis. Remember, that in EEG, one often has to deal with locally bad sensors. And in this context, it is noteworthy that none of the other methods we have discussed so far provides an explicit treatment for single trial analysis in the presence of locally bad sensors. Our comparison to the RANSAC algorithm seems to suggest that the RANSAC algorithm is indeed sensitive to the parameter settings. Even though the default settings appear to work reasonably well for the EEG data (Figure~\ref{fig:dgw_scatter}), they are not so optimal for the HCP MEG data (Figure~\ref{fig:hcp_scatter}).

It is perhaps worth emphasizing that using cross-validation implies that the trials with artifacts are independent. If this assumption is violated and if artifacts are phase-locked between the training and validation sets, \emph{i.e.} occur for all trials at the same time relative to trial onsets, then this can interfere with the estimation procedure in \emph{autoreject}. Another caveat to be noted is that if the data contains more than $\rho^{*}$ (the maximum number of sensors that can be interpolated) bad sensors, and if the trial is not dropped, the data in the remaining bad sensors can still spread to other sensors if one were to use spatial filters such as SSP. Finally, \emph{autoreject} considers only peak-to-peak thresholds for detecting bad sensors. Of course, the user must still mark low-amplitude flat sensors using another threshold; however, a simple threshold would suffice here as such sensors are usually consistently flat.
Regardless of the method that researchers choose to adopt, diagnostic plots and automated reports~\citep{dengemann2015conc} are an essential element to assess and better understand possible failures of automatic procedures. In this regard, transparency of the method in question is important. In the case of our \emph{autoreject (local)} implementation, we offer the possibility for the user to inspect the bad segments marked by the automated algorithm and correct it if necessary. An example of such a plot is shown in Figure~\ref{fig:diagnostic_plot}. Automating the detection of bad sensors and trials has the benefit of avoiding any unintentional biases that might be introduced if the experimenter were to mark the segments manually. In this sense, diagnostic visualization should supplement the analysis by ensuring accountability in the case of unexpected results.

\subsection{Autoreject in the context of ICA, SSP and SSS}

It is now important to place these results in the broader context of electrophysiological data analysis. Regarding the correction of specific artifacts such as electrooculogram (EOG) artifacts, \emph{autoreject (local)} does indeed remove or interpolate some of the trials affected by eye blinks. This is because most eye blinks are not time-locked to the trial onsets and therefore get detected in the cross-validation procedure. However, the weaker eye blinks, particularly those smaller in magnitude than the evoked response, are not always removed. Also, the idea of rejection is to remove extreme values which are supposed to be rare events. This is why our empirical observation suggests that \emph{autoreject (local)} is not enough in the presence of too frequent eye blinks, but also not enough to fully get rid of the smallest EOG artifacts. 

This is where ICA~\citep{vigario1997extraction} and Signal Space Projection (SSP)~\citep{uusitalo1997signal} can naturally supplement \emph{autoreject}. These methods are usually employed to extract and subsequently project out signal subspaces governed by physiological artifacts such as muscular, cardiac and ocular artifacts. However the estimation of these subspaces can be easily corrupted by other even more dramatic environmental or device-related artifacts. This is commonly prevented by band-pass filtering the signals and excluding high-amplitude artifacts during the estimation of the subspaces. Both ICA and SSP (particularly if it estimated from the data rather than an empty room recording) are highly sensitive to observations with high variance. Even though they involve estimating spatial filters that do not incorporate any notion of time, artifacts very localized in time will very likely have a considerable impact on the estimation procedure. This leads us to recommend removing globally bad sensors and employing appropriate rejection thresholds to exclude time segments with strong artifacts. 

The success of applying \emph{autoreject} to any electrophysiological data hinges critically on its ability to isolate artifacts local in time which cannot necessarily be identified by a prototypical spatial signature. However, the spatial interpolation employed by \emph{autoreject} may not be able to repair sensors which are clustered together. In this case, the software package that implements the spatial interpolation should warn the user if the error due to the interpolation is likely to be high. Such a cluster of bad sensors can be expected in the case of physiological artifacts, such as muscular, cardiac or ocular artifacts. To take care of such artifacts with prototypical spatial patterns, ICA is certainly a powerful method, yet manual identification of artifactual components remains today done primarily manually.

If the context of data processing supports estimation of ICA and SSP on segmented data, we would recommend to perform it after applying \emph{autoreject}, benefiting from its automated bad sensor and bad trial handling. MEG signals usually contain a strong contribution from environmental electromagnetic fields. Therefore, interference suppression of MEG data is often needed, utilizing hardware and software based approaches (see, \textit{e.g.} \citet{parkkonen:2010} for details). In principle, spatial interpolation of bad sensor signals may not work very well unless the environmental interference has been removed. In the present study, the MNE sample data was recorded in a very well shielded room and did not need separate interference suppression, while the interference in the 4D/BTi data was removed by utilizing the reference channels. Spatial filtering approaches, such as SSP or SSS, may however produce a ``chicken and egg'' dilemma -- whether to apply SSP/SSS or autoreject first - which can be solved using an iterative procedure as suggested by the PREP pipeline~\citep{bigdely2015prep}. That is, first run autoreject only for detection of bad channels but without interpolation. This is followed by an SSS run excluding the bad channels detected by autoreject. Finally, autoreject can be applied on the data free of environmental interference.

\subsection{Source localization with artifact rejection}

Obviously, artifact-free data benefits almost any analysis that is subsequently performed and the M/EEG inverse problem is no exception. Such benefits not only concern the quality of source estimates but also the choice of source-localization methods, as some of these methods require modification when certain artifact rejection strategies are employed.
As \emph{autoreject} amounts to automating a common, pre-existing and early processing step it does not require changes for source-level analyses. For example, evoked responses obtained using \emph{autoreject (local)} can be readily analyzed with various source localization methods such as beamformer methods~\citep{dalal2008five,gross2001dynamic}, or cortically-constrained Minimum Norm Estimates with $\ell_2$ penalty~\citep{uutela1999visualization}, and noise-normalized schemes, such as dSPM~\citep{dale2000dynamic} and sLORETA~\citep{pascual2002standardized}. 

Certain denoising techniques such as SSP~\citep{uusitalo1997signal} or SSS~\citep{taulu2004suppression} reduce the rank of the data which can be problematic for beamforming techniques~\citep{woolrich2011meg}. This needs special attention, and in some software such as MNE, this is handled using a non-square whitening matrix. However, as \emph{autoreject} does not systematically reduce the rank of the data, it does not even require sophisticated handling of the data rank. At the same time, it works seamlessly with noise-normalization, where the estimation of the between-sensor noise covariance depends on the number of trials. To estimate the noise covariance during baseline periods, one computes the covariance of non-averaged data and then, assuming independence of each trial, the covariance gets divided by the number of trials present in the average~\citep{engemann2015automated_new}. Most existing pipelines scale the covariance by an integer number of trials. In contrast, methods such as robust regression~\citep{diedrichsen2005detecting} that preferentially give less weight to noisy trials, require the noise normalization to be modified. Concretely, one would have to estimate an approximate number of trials or estimate the covariance matrix by restricting the computation to a subset of trials. \emph{Autoreject} does not necessitate any such modifications to the source-localization pipeline, and hence, helps reduce the cognitive load of integration with pre-existing tools.

%% file: sections/conclusion.tex
\section{Conclusion}

In summary, we have presented a novel algorithm for automatic data-driven detection and repair of bad segments in single trial M/EEG data. We therefore termed it \emph{autoreject}. We have compared our method to state-of-the-art methods on four different open datasets containing in total more than 200 subjects. Our validation suggests that \emph{autoreject} performs at least as good as diverse alternatives and commonly used procedures while often performing considerably better. This is the consequence of the combination of a data-driven outlier-detection step combined with physics-driven channel repair where all parameters are calibrated using a cross-validation strategy robust to outliers. The insight about the necessity to tune parameters at the level of single sensors and for individual subjects was further consolidated by our analyses of threshold distributions. The empirical variability of optimal thresholds across datasets emphasizes the importance of statistical learning approaches and automatic model selection strategies for preprocessing M/EEG signals. While \emph{autoreject} makes use of advanced statistical learning techniques such as Bayesian hyperparameter optimization, it is also grounded in the physics underlying the data generation. It is therefore not purely a black-box data-driven approach. It balances the trade-off between accuracy and interpretability. Indeed all \emph{autoreject} parameters have a meaning from a user standpoint and the algorithmic decisions can be explained. Supplemented by efficient diagnostic visualization routines, \emph{autoreject} can be easily integrated in MEG/EEG analysis pipelines,
including clinical ones where understanding algorithmic decisions is mandatory for tool adoption.

By offering an automatic and data-driven algorithmic solution to a task mostly so far done manually, \emph{autoreject} reduces the cost of data inspection by experts. By allowing to repair data rather than removing it from the study, it allows saving data which are also costly to acquire. In addition, it removes the experts' bias which are due to specific training or prior experience, as well as some expectations about the data. It does so by defining a clear set of rules serving as inclusion criteria for M/EEG data, making results more easily reproducible and eventually limiting the risk of false discoveries. Furthermore, as data sharing across centers has become a common practice, \emph{autoreject}  addresses the issue of heterogeneous acquisition setups. Indeed, each acquisition set-up has its intrinsic signal qualities, which means that preprocessing parameters can vary significantly between datasets. As opposed to alternative methods, \emph{autoreject} automates the estimation of its parameters.

%% file: sections/acknowledgement.tex
\section{Acknowledgement}

We thank Lionel Naccache for providing us with dramatic examples of artifact-ridden clinical EEG data which considerably stimulated the research presented in this study. The work was supported by the French National Research Agency (ANR-14-NEUC-0002-01), the National Institutes of Health (R01 MH106174) and ERC Starting Grant SLAB ERC-YStG-676943. Denis A. Engemann acknowledges support by the Amazon Webservices Research Grant awarded to him and the ERC StG 263584 awarded to Virginie van Wassenhove. We thank the Cognitive Neuroimaging Unit at Neurospin for fruitful discussions and feedback on this work. We further thank the MNE-Python developers and the MNE community for continuous collaborative interaction on basic development of the academic MNE software without which this study could not have been conducted.

%% file: sections/supplementary.tex
\newpage
\renewcommand{\thefigure}{S\arabic{figure}}
\setcounter{figure}{0}

\section*{Supplementary material}

Here, we present additional material that may answer some questions that the reader might have when reading the main text.

\begin{figure}[htb]
	\centering
	\includegraphics[width=0.9\linewidth]{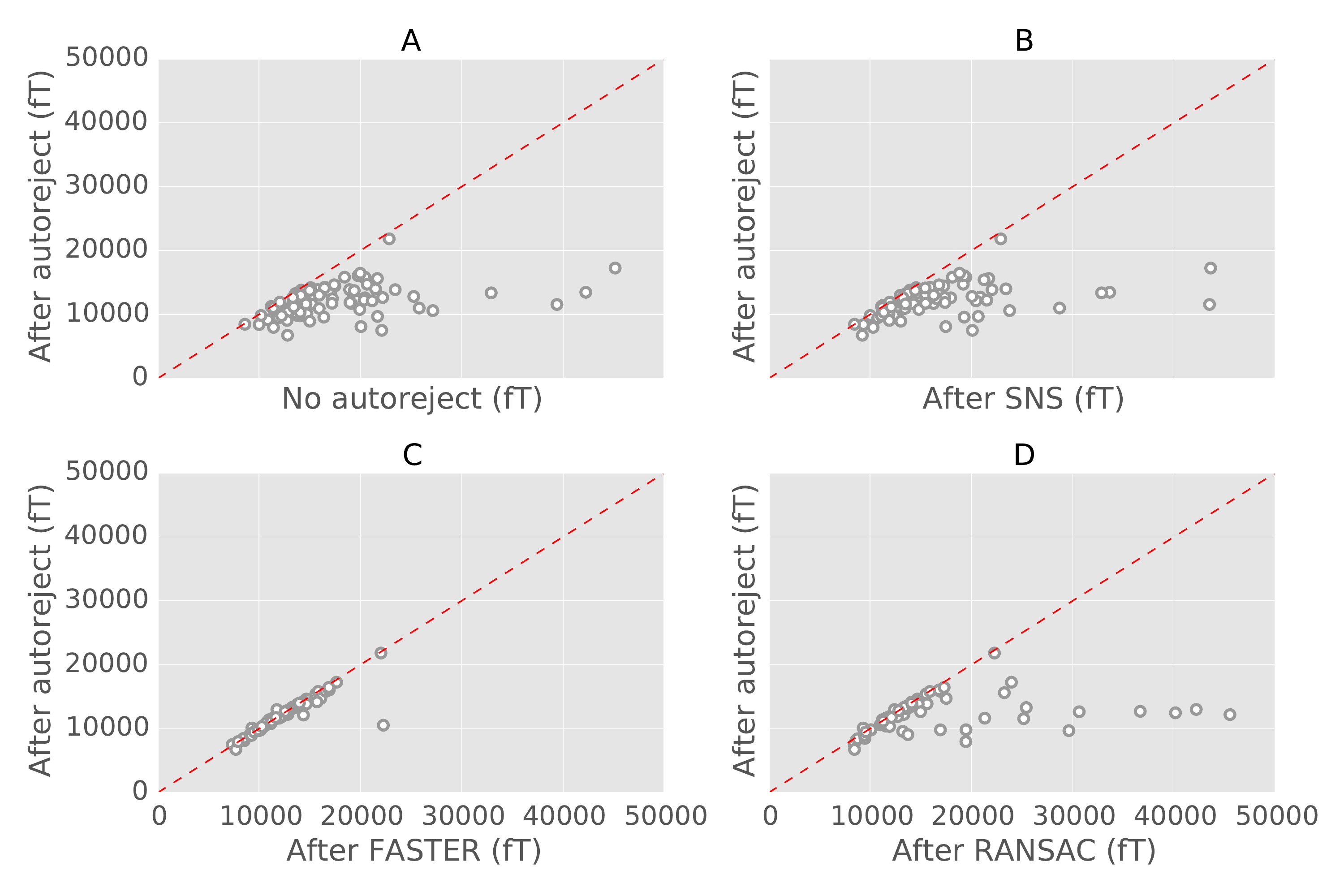}
    \caption{Scatter plots for the results with the HCP data. This figure uses the same data as in Figure~\ref{fig:hcp_scatter} from the main text, but with  $\|\cdot\|_2$ norm instead of the $\|\cdot\|_\infty$ norm for computing the difference between the HCP ground truth and the method. As before, each circle is a subject. (A) \textit{autoreject (local)} against no rejection, (B) \textit{autoreject (local)} against Sensor Noise Suppression (SNS) (SNS), (C) \textit{autoreject} against FASTER, (D) \textit{autoreject (local)} against RANSAC. Data points below the dotted red line indicate subjects for which \textit{autoreject (local)} outperforms the alternative method.}
    \label{fig:l2_norm}
\end{figure}

\paragraph{$\ell_{2}$ vs $\ell_{\infty}$ norm}: Why not use $\ell_{2}$ norm instead of $\ell_{\infty}$ norm to report the quantitative results in Figure~\ref{fig:hcp_scatter} or Figure~\ref{fig:dgw_scatter}? The reason is that the $\ell_{2}$ norm will average across the sensors. If one sensor is badly corrupted, then this would not be obvious with the $\ell_{2}$ norm because the average in the $\ell_{2}$ norm computation conceals the isolated problematic sensors with large artifacts. However, as the $\ell_\infty$ norm captures the worst sensor, it can be used to visualize pathological cases where even one sensor is corrupted. In Figure~\ref{fig:l2_norm}, we reproduce Figure~\ref{fig:hcp_scatter} using the $\ell_2$ norm instead of $\ell_{\infty}$. We can observe that, although the pattern remains the same, it is much less clear where one method outperforms the other. Even where \emph{autoreject} isn't performing as well, it is not  visible due to the averaging effect.

